\documentclass[conference]{IEEEtran}
\IEEEoverridecommandlockouts

\usepackage{cite}
\usepackage{amsmath,amssymb,amsfonts}
\usepackage{algorithmic}
\usepackage{graphicx}
\usepackage{textcomp}
\usepackage{xcolor}
\usepackage{booktabs}
\usepackage{subcaption}
\usepackage{url}
\usepackage{hyperref}
\usepackage{fancyhdr}
\def\BibTeX{{\rm B\kern-.05em{\sc i\kern-.025em b}\kern-.08em
    T\kern-.1667em\lower.7ex\hbox{E}\kern-.125emX}}

\def\BibTeX{{\rm B\kern-.05em{\sc i\kern-.025em b}\kern-.08em
    T\kern-.1667em\lower.7ex\hbox{E}\kern-.125emX}}
\renewcommand{\baselinestretch}{0.94}
\begin{document}




\title{AdaptViT: Runtime-Adaptive Vision Transformer Deployment on Custom RISC-V}

\author{
\IEEEauthorblockN{
Vishnu PS,
Ajay Kumar M,
Yike Li,
Robert Bogdan Staszewski,
Deepu John
\thanks{This work was supported by Taighde Éireann - Research Ireland (22/US/3848, 21/RP-2TF/10019).}
}

\IEEEauthorblockA{University College Dublin, Ireland}
}

\maketitle

\begin{abstract}
Deploying Vision Transformers (ViTs) on low-power edge devices is challenging due to high computational demands. Conventional pruning frameworks require a separate compiled binary for each sparsity level, increasing storage overhead and limiting runtime adaptability. This paper presents an end-to-end deployment pipeline that transforms pretrained ViTs into a single runtime-configurable binary, enabling dynamic compute-budget switching on embedded CPUs. This is achieved by restructuring generated C kernels with modified loop bounds and binary-mask control logic, allowing execution to switch across discrete sparsity levels via compact external configuration files. Compared to multi-binary deployment, the proposed runtime-adaptive approach reduces on-device storage by up to $4.86\times$, requiring only 163 MB for ViT-Base instead of nearly 800 MB. To maximize pruning efficiency, we introduce a hardware-aligned block pruning strategy for Multi-Layer Perceptron (MLP) layers. In addition, a custom ISA extension is proposed to exploit input-reuse patterns in linear projection kernels. On a Synopsys TRV32P3FX RISC-V processor, the full system achieves up to $2.8\times$ speedup at 65\% MLP and 50\% attention-head pruning for ViT-Base. The ISA extension alone provides a $1.56\times$ speedup and 33\% lower inference energy, with a 24.7\% area overhead in a TSMC 28~nm implementation.
\end{abstract}

\vspace{-2mm}
\section{Introduction}

Deploying deep learning models on edge devices requires high computational efficiency without significant loss in accuracy \cite{liu2022bringing}. However, embedded processors are constrained by limited memory and compute resources. Vision Transformers (ViTs) achieve strong performance but incur high computational and memory costs due to Multi-Head Self-Attention (MHSA) and large Feed-Forward Networks (FFNs), limiting their suitability for embedded deployment \cite{dosovitskiy2020image, kaneko2024analysis, chen2024comprehensive}. Model pruning is therefore widely used to reduce complexity by trading accuracy for efficiency. However, edge environments are dynamic, with changing constraints due to battery levels, workload changes, or user demands. This motivates the need for multiple operating points with runtime adaptability and minimal overhead. A straightforward solution of storing and switching between multiple pruned binaries is impractical on microcontrollers due to high flash usage and the need to reload models when switching configurations.

Existing approaches can be broadly categorized into dynamic and static methods. Dynamic methods adapt computation at runtime. For instance, token pruning in ViTs \cite{rao2021dynamicvit, dong2023heatvit} removes uninformative tokens per input to reduce attention cost, but introduces irregular execution and memory access patterns that are difficult to optimize on embedded hardware. Static structured pruning methods, such as LPViT \cite{xu2024lpvit}, apply hardware-aware block pruning to improve efficiency but result in fixed sparsity configurations, limiting runtime switching across operating points. Parikh et al. \cite{parikh2024accelerating} combine weight pruning with input-adaptive token pruning on FPGA accelerators; however, the weight pruning stage requires retraining and yields a fixed sparsity structure, so each operating point still requires a separately trained model. Once-for-All \cite{cai2019once} enables multi-width inference within a single network, but relies on specialized training from scratch and cannot directly leverage pretrained models. SnapViT \cite{simoncini2025elastic} enables multi-level post-training pruning of pretrained ViTs using importance and correlation scores, generating multiple sparsity configurations without retraining. However, evaluation is limited to Python simulation and does not address embedded deployment, where compiler toolchains generate static C execution paths that make runtime-adaptive pruning difficult to realize efficiently. To address this gap, we propose an end-to-end deployment pipeline that enables runtime-adaptive ViT inference on RISC-V processors starting from pretrained models. Our main contributions are:

\begin{itemize}
\item \textbf{Hardware-aware structured static pruning for ViTs:} We extend the SnapViT pruning framework \cite{simoncini2025elastic} with block-structured MLP pruning, replacing neuron-level pruning to improve hardware efficiency.
\item \textbf{Runtime-configurable inference pipeline:} We propose an end-to-end framework that enables runtime-adaptive inference of pretrained ViT on RISC-V processors from a single compiled binary, with minimal overhead.
\item \textbf{ISA-level acceleration:} We introduce a custom MAC2 instruction to RISC-V ISA to further accelerate ViT inference.
\end{itemize}
\vspace{-4mm}
\section{Methodology}

\begin{figure}[t]
    \centering
    \includegraphics[width=0.7\columnwidth]{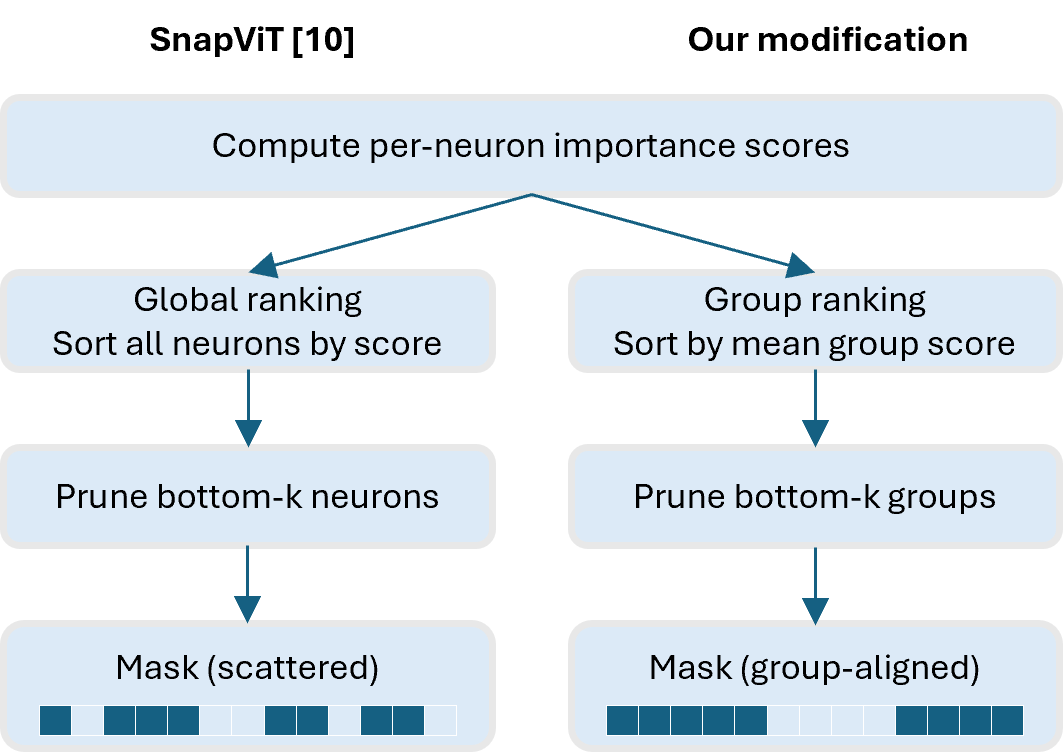}
    \caption{MLP Pruning Method.}
    \label{fig:alg_flow}
    \vspace{-2mm}
\end{figure}

Pretrained ViTs are converted into a runtime-adaptive embedded binary through a set of offline processing stages. First, post-training pruning is applied to generate multiple sparsity levels, extending \cite{simoncini2025elastic} with hardware-aligned block pruning for MLP layers while retaining per-head pruning for attention blocks. As shown in Fig.~\ref{fig:alg_flow}, neurons are grouped into blocks of eight and ranked using mean importance before pruning at the block level. To recover accuracy after pruning, a task-specific linear classifier is trained via linear probing for each sparsity level while keeping the transformer backbone frozen. The unpruned backbone is then partially quantized to INT8 using \textit{TORCHAO} \cite{or2025torchao}, where linear projection layers are quantized while nonlinear operations remain in floating point to preserve accuracy. The resulting model is compiled to bare-metal C using TVM and post-processed to add mask-based skip logic and loop restructuring (Section~\ref{subsection:Post-Compilation C Code Transformation}), enabling runtime selection of sparsity levels within a single binary. The final implementation is deployed on RISC-V core, where the pruning level selected at run time loads the corresponding pruning mask and classifier weights.

\vspace{-2mm}
\subsection{C Code Transformation}
\label{subsection:Post-Compilation C Code Transformation}

While dynamic selection of active dimensions is straightforward in high-level environments such as Python, reproducing this behavior in bare-metal C is challenging. During edge deployment, high-level runtime abstractions are removed and TVM generates optimized C code with fixed memory layouts and static loop bounds. Introducing conditional mask checks inside compute-intensive regions adds control-flow overhead, reducing efficiency. Therefore, exploiting sparsity on embedded targets requires a post-compilation transformation that restructures loops to enable safe skipping of computation at coarse-grained block boundaries with minimal overhead.

The TVM compiler maps ViT operations into C kernels across encoder blocks. In the generated code, linear projections (QKV projection, output projection, FC1, and FC2) are implemented as four \texttt{dense\_pack} kernel variants, repeated across 12 transformer blocks. Similarly, attention score computation and value aggregation are implemented using two \texttt{fused\_nn\_batch\_matmul} kernels. Profiling on the target RISC-V processor shows these kernels dominate execution, accounting for approximately 79\% and 83\% of total inference cycles for ViT-Small and ViT-Base, respectively. Since structured pruning targets attention heads and MLP blocks, optimizing only these kernels captures most achievable speedup. We use AugReg ViT-Small ($d=384$, 6 heads) as the reference model for the remainder of this section.

\subsubsection{Dense Kernel Restructuring}

All four \texttt{dense\_pack} kernels share the same inner-loop structure, where a single input activation is multiplied with eight weights and accumulated into eight outputs per iteration,  defining an 8-output compute granularity. However, the dimension along which sparsity appears differs across kernel types. Table~\ref{tab:dense_layouts} summarizes tensor layouts across ViT-Small. In QKV projections, the 1152 output channels correspond to concatenated $Q$, $K$, and $V$, with each attention head spanning 64 channels. A pruned head therefore removes its associated $Q/K/V$ slices. Although TVM tiles this GEMM into 128-column blocks (two heads per tile), we restructure execution to operate at head granularity, enabling head-wise skipping and saving $197 \times 384 \times 64 \times 3$ MACs per pruned head. In the output projection, all heads get fused into a 384-dimensional representation, so each output column depends on all heads and tile-level skipping is not possible. Instead, we decompose the 384-dimension dot product into six 64-dimensional head-aligned segments and skip pruned heads, eliminating $197 \times 64 \times 384$ MACs per head. For the MLP (FC1/FC2), FC1 expands $d=384$ to $1536$ and FC2 projects back. TVM splits FC1 into 3 tiles of 512 columns, further decomposed into 64 groups of 8 MACs matching the inner-loop width. In neuron-level pruning (SnapViT), sparsity is irregular, so even at 65\% sparsity only $328/2304$ eight-wide MAC groups are fully inactive, forcing execution of most 8-MAC groups and yielding only 14.2\% MAC reduction. In contrast, enforcing group-of-8 structured pruning aligns sparsity with hardware granularity, allowing full group skipping and achieving 65.0\% MAC reduction (4.6$\times$ improvement). Since FC1 and FC2 share the same mask, skipped groups propagate across both layers, compounding savings.

\begin{table}[t]
\centering
\caption{Dense and Batch Kernel Tensor Layouts}
\label{tab:dense_layouts}
\setlength{\tabcolsep}{2.2pt} 
\renewcommand{\arraystretch}{0.7}
{\small
\begin{tabular}{lccc}
\toprule
\textbf{Layer} & \textbf{Input $X$} & \textbf{Weight/Input $W$} & \textbf{Output $Y$} \\
\midrule
QKV Proj.   & $\mathbb{R}^{197 \times 384}$ & $\mathbb{R}^{384 \times 1152}$ & $\mathbb{R}^{197 \times 1152}$ \\ \addlinespace
Attention 1 & $\mathbb{R}^{6 \times 197 \times 64}$  & $\mathbb{R}^{6 \times 64 \times 197}$  & $\mathbb{R}^{6 \times 197 \times 197}$  \\ \addlinespace
Attention 2 & $\mathbb{R}^{6 \times 197 \times 197}$ & $\mathbb{R}^{6 \times 197 \times 64}$  & $\mathbb{R}^{6 \times 197 \times 64}$   \\ \addlinespace
Out Proj.   & $\mathbb{R}^{197 \times 384}$ & $\mathbb{R}^{384 \times 384}$  & $\mathbb{R}^{197 \times 384}$  \\ \addlinespace
FC1 (MLP)   & $\mathbb{R}^{197 \times 384}$ & $\mathbb{R}^{384 \times 1536}$ & $\mathbb{R}^{197 \times 1536}$ \\ \addlinespace
FC2 (MLP)   & $\mathbb{R}^{197 \times 1536}$ & $\mathbb{R}^{1536 \times 384}$ & $\mathbb{R}^{197 \times 384}$  \\
\bottomrule
\end{tabular}
}
\vspace{-3mm}
\end{table}



\subsubsection{Batch Matmul Restructuring}

This class of kernels performs attention score computation ($\mathbf{Q}\mathbf{K}^\top$) and value aggregation ($\mathbf{A}\mathbf{V}$), offering a direct opportunity to exploit head-level sparsity. Since pruned heads produce zero-valued $\mathbf{Q}$, $\mathbf{K}$, and $\mathbf{V}$ in the preceding stage, their corresponding outputs in both kernels are also zero. We exploit this by introducing a binary head mask at kernel call. Active heads are executed via range-restricted calls, while pruned heads are skipped and their output regions are zero-initialized using a single \texttt{memset}. This preserves exact functional equivalence while eliminating $197 \times 197 \times 64$ MAC operations per pruned head per kernel.
\vspace{-2mm}

\subsection{Framework Generalization}

Although ViT variants differ in dimensions, the TVM-generated kernel structure and execution order remain consistent. This enables the same post-compilation transformations to be applied across models ViT-Small and ViT-Base. We automate this process via a Python script that extracts loop bounds, tile sizes, and memory offsets for each architecture.
\vspace{-2.5mm}
\subsection{ISA Extension}
\label{subsection:ISA Extension}

The innermost loop of all four \texttt{dense\_pack} kernels performs eight MAC operations per iteration, reusing a single input activation across eight weights. While the baseline TRV32P3FX core supports fused floating-point MACs via the \texttt{F} extension, the quantized workload is dominated by integer operations, requiring 16 instructions (8 \texttt{mul} + 8 \texttt{add}) per iteration. A direct multi-MAC instruction is limited by the five-register operand constraint of the 32-bit RISC-V encoding. However, the input-reuse pattern in the kernel enables a dual-MAC instruction operating on a shared activation $a$, two weights $b_0,b_1$, and two accumulators $c_0,c_1$:
\begin{equation}
\texttt{MAC2}(c_0,c_1,a,b_0,b_1) \Rightarrow
c_0 \mathrel{+}= a b_0,\quad
c_1 \mathrel{+}= a b_1
\end{equation}
The custom hardware instantiates two parallel MAC pipelines sharing operand $a$, broadcasting it to both multipliers while independently processing $b_0$ and $b_1$, and writing back $c_0$ and $c_1$ in parallel. This reduces the innermost loop from 16 instructions to 4. While general-purpose vector (RVV) or packed-SIMD (RVP) extensions could offer higher flexibility, they introduce significant area and architectural complexity. In contrast, the proposed \texttt{MAC2} provides SIMD-like acceleration tailored to dominant projection kernels with substantially lower hardware overhead.



            \begin{figure}
        \centering
        \includegraphics[width=0.9\columnwidth]{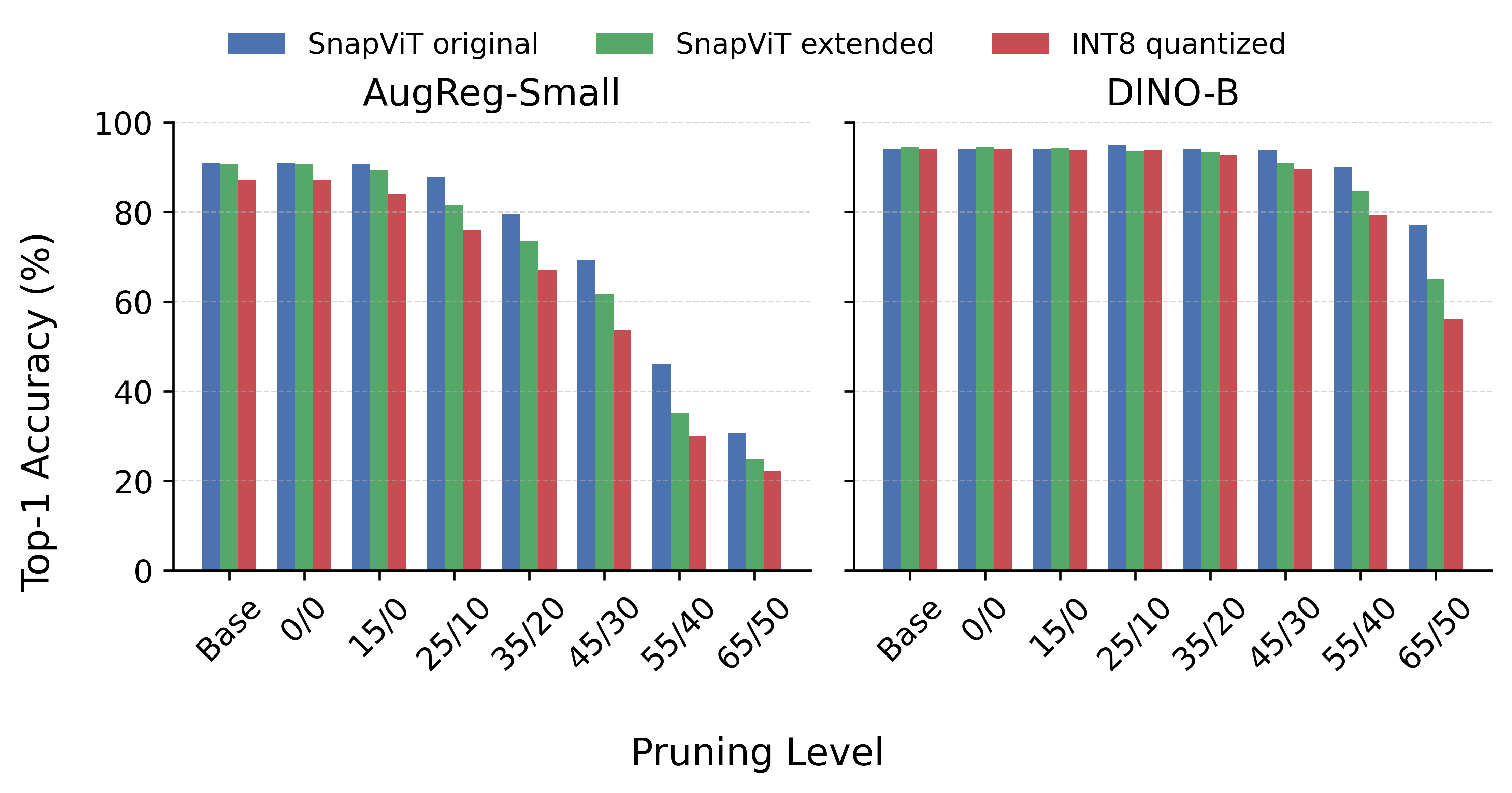}
        \caption{Accuracy vs.\ pruning level}
        \label{fig:accuracy_graph}
        \vspace{-2mm}
    \end{figure}


    \begin{figure}[t]
        \centering
        \includegraphics[width=0.78\columnwidth]{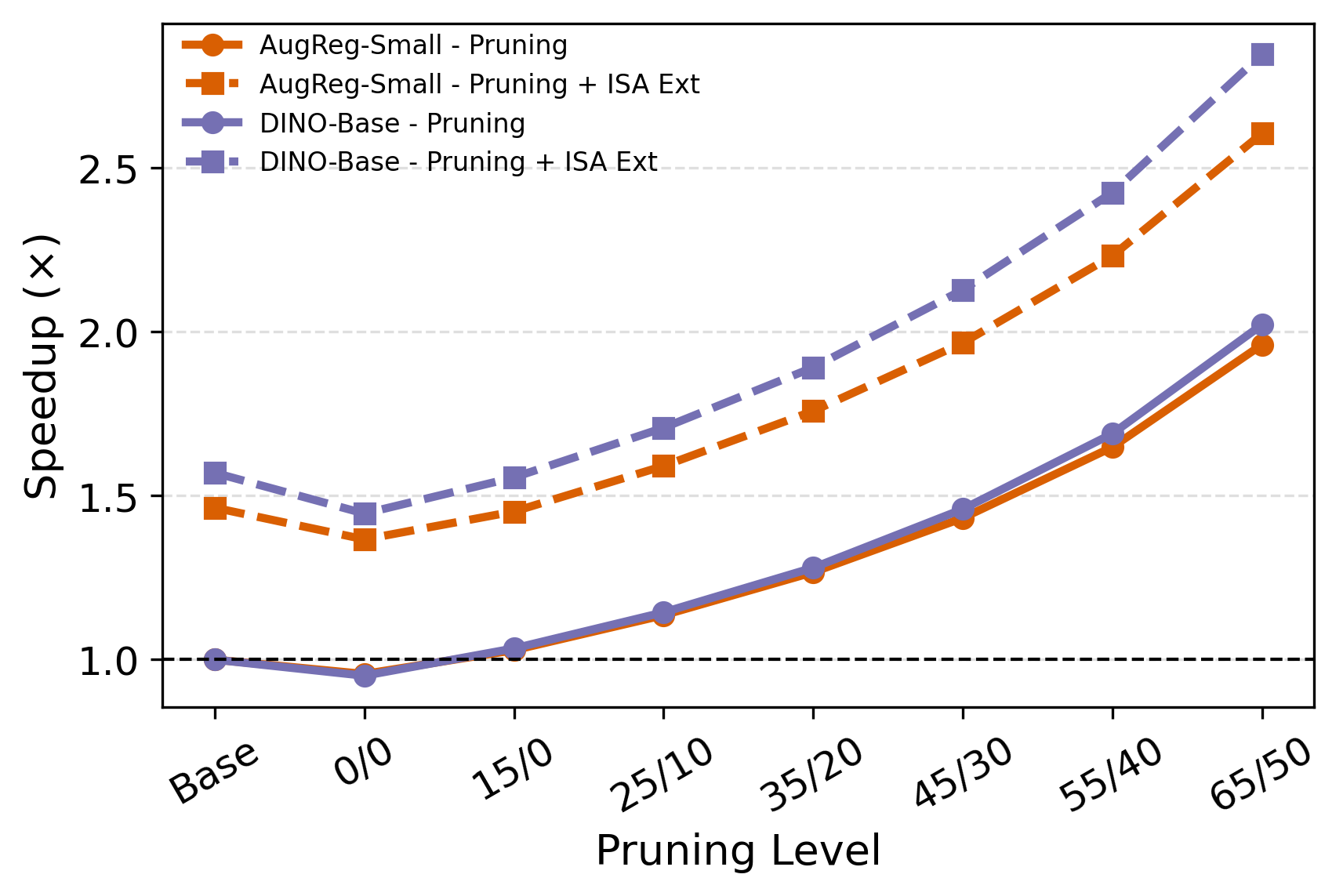}
        \caption{ Acceleration with pruning and ISA extension}
        \label{fig:inf_accn}
        \vspace{-5mm}
    \end{figure}

    \begin{figure}[t]
        \centering
        \includegraphics[width=0.78\columnwidth]{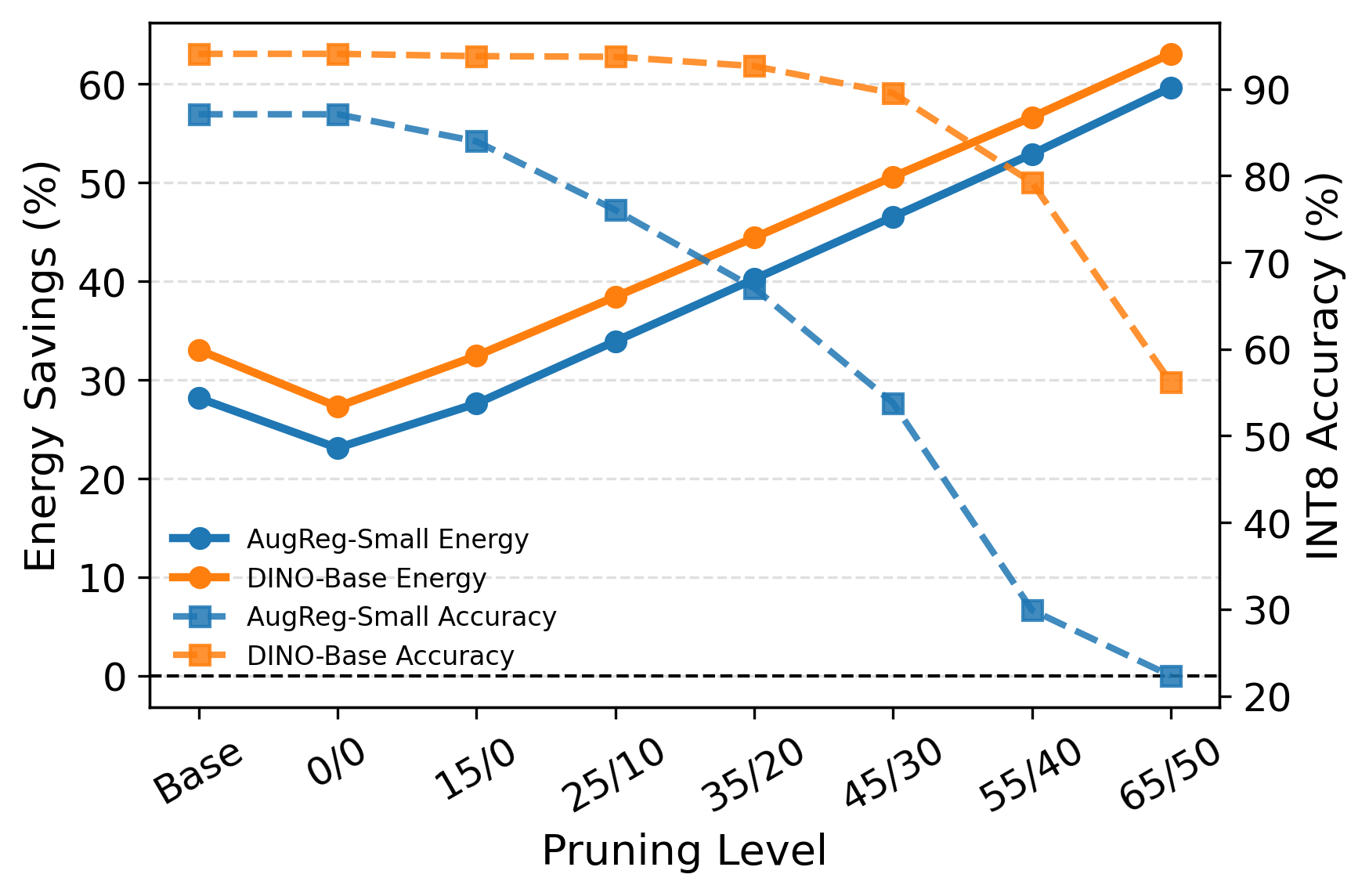}
        \caption{ Trade-off between energy efficiency and INT8 accuracy.}
        \label{fig:energy_acc_trafeoff}
        \vspace{-6mm}
    \end{figure}

\vspace{-3.5mm}
\section{Results}

Pruning mask generation across multiple sparsity levels, quantization, and C code generation using TVM (v0.14.dev273) are performed offline on an NVIDIA A100 GPU in Google Colab. Evaluation is conducted on the Oxford-IIIT Pet dataset (37 classes). For each pruning configuration, linear probing is performed, with backbone frozen and top-1 accuracy reported on the test set. The generated C code is post-processed locally and compiled for the TRV32P3FX RISC-V core using Synopsys ASIP Designer (W-2024.12). TRV32P3FX is a 32-bit, 3-stage in-order RISC-V core with single-precision floating-point support, 2-way instruction-level parallelism, and hardware loops. Compilation uses -O3, and latency is measured via instruction-accurate (IA) simulation on both base and ISA-extended cores. RISC-V results are validated against PyTorch quantized inference, confirming identical top-1 predictions. The implementation code will be released as a GitHub repository\footnote{\url{https://github.com/psvishnu1/AdaptViT}} to support reproducibility.
\vspace{-1mm}

\subsection{Accuracy Impact}

Fig.~\ref{fig:accuracy_graph} shows accuracy across six pruning configurations for two ViT variants. The x-axis denotes runtime settings as $X/Y$, where $X$ and $Y$ represent MLP group and attention head pruning ratios, respectively. Each configuration is evaluated under three execution modes. The first corresponds to FP32 per-neuron pruning. The second is FP32 inference with our group-of-8 structured pruning, introducing hardware-aligned sparsity, but resulting in an average accuracy drop of 2.12\% over baseline. The third applies INT8 quantization, leading to an additional 5.07\% drop relative to FP32. At the cost of this small degradation, INT8 significantly reduces memory and compute cost, making it suitable for embedded deployment. 

\vspace{-1mm}
\subsection{Storage Requirement Analysis}

Let $d$, $d_mlp$, $L$, and $b$ denote the embedding dimension, MLP dimension, number of transformer blocks, and bytes per weight. Each block contains attention parameters ($A = 4d^2$) and MLP parameters ($M = 2.d.d_{mlp}$) with total size $T = 12d^2$ (using $d_{mlp}=4d$ and fixed split $A/T = 1/3$, $M/T = 2/3$). For pruning configuration $i$, with MLP-pruning ratio $r_{\text{mlp}}^{(i)}$ and head-pruning ratio $r_{\text{head}}^{(i)}$, the retained weight fraction $f_i$ and corresponding storage $W_i$ are: $f_i = \tfrac{1}{3}(1 - r_{\text{head}}^{(i)}) + \tfrac{2}{3}(1 - r_{\text{mlp}}^{(i)}), \; W_i = 12Ld^2 b f_i.$ In the conventional approach, each pruning level is stored as a separate model, leading to total storage $S_{\text{separate}} = \sum_{i=1}^{N} W_i$ for N pruning levels, plus a small code-size overhead due to separate kernel layouts. In the proposed method, a single full model $W_{\text{full}} = 12Ld^2 b$ is stored once and shared across all configurations. Runtime adaptability is enabled via lightweight per-variant metadata: (i) pruning masks and (ii) a classifier head. With $H$ denoting the number of attention heads per block, the MLP and head masks contain $4Ld$ and $LH$ entries respectively; stored in FP32, this gives mask overhead $4L(4d + H)$. Each variant also stores a classifier consisting of a $d \times K$ INT16 weight matrix and a $K$-dimensional FP32 bias, where $K$ is the number of output classes for the downstream task, giving $dKb + 4K$. Hence, for $N$ configurations: $S_{\text{proposed}} = W_{\text{full}} + N\big[4L(4d+H) + dKb + 4K\big].$ Overall, across ViT-Small and ViT-Base, the proposed approach reduces on-device storage by up to $4.86\times$ over separate per-pruning-level binaries, requiring only 41.18 MB vs 198.45 MB and 163.37 MB vs 793.80 MB respectively.

\subsection{Runtime Switching Overhead}


Switching between pruning configurations incurs minimal overhead. The same executable and shared weight tensor are reused across all operating points. Pruning masks are passed at runtime as references to external stored buffers, and kernel-level skip logic uses these masks to determine active computation. Thus, switching configurations reduces to redirecting pointers to the corresponding mask buffers. The only remaining overhead is classifier handling, where the weight matrix and bias are currently copied into a fixed memory location during switching. This cost is negligible compared to full model reload or reinitialization and does not affect real-time execution. The switching costs ${\sim}57$K / $114$K cycles for ViT-Small / Base, respectively, accounting for less than $10^{-5}\times$ of single inference cycles. It can be fully eliminated by storing classifier heads as pre-allocated buffers and accessing them via pointers, consistent with the mask mechanism; this is left as future work.

\subsection{Inference Speedup Evaluation}

Fig.~\ref{fig:inf_accn} shows speedup across pruning levels for ViT-Small and ViT-Base (DINO),  across two settings. The first captures pruning-only speedup on the baseline RISC-V core relative to the unpruned model. The second includes both pruning and the proposed ISA extension, normalized to the unpruned model on the un-extended core. An all-ones mask ($0/0$) run isolates runtime control overhead, which remains below 5\%, confirming negligible skip-logic cost. The \texttt{MAC2} instruction provides up to 1.56$\times$ speedup for the unpruned model, leading to a combined speedup of up to 2.8$\times$ at the highest pruning level for ViT-Base. Fig.~\ref{fig:energy_acc_trafeoff} presents energy and accuracy trends across pruning levels on the extended RISC-V, relative to the unpruned baseline. Energy per inference is calculated as $E = P \cdot C/f$, where $P$ denotes ASIC synthesis estimated power, $C$ the IA simulation cycle count, and $f$ the operating frequency. At zero pruning, the observed 33\% energy reduction is solely due to the \texttt{MAC2} extension. Additional pruning progressively reduces energy consumption due to lower MAC operations. The INT8 accuracy curve shows the corresponding drop under increasing sparsity. Together, these trends highlight the energy–accuracy trade-off, allowing selection of operating points based on application constraints. ViT-Small shows larger accuracy degradation due to lower redundancy compared to ViT-Base.
\subsection{Hardware Overhead}
\begin{table}[t]
\centering
\caption{Hardware Overhead of the Extended Processor}
\label{tab:implementation_overhead}
\setlength{\tabcolsep}{3.5pt}
\renewcommand{\arraystretch}{0.95}

\begin{tabular}{lcc}
\toprule
\textbf{Metric} & \textbf{Base} & \textbf{Extended} \\
\midrule

\multicolumn{3}{c}{\textit{FPGA Implementation (ZCU102 @ 65 MHz)}} \\
\midrule

LUT / FF       & 10570 / 2186 & 14931 / 2236 \\
DSP / Power    & 6 / 0.186\,W  & 12 / 0.208\,W \\
\midrule

\multicolumn{3}{c}{\textit{ASIC Synthesis (TSMC 28 nm @ 100 MHz)}} \\
\midrule

Area ($\mu m^2$)      & 21104   & 26315 \\
Power (mW)            & 0.3901  & 0.4100 \\
Normalized ADP        & 1.00    & 0.79 \\
\bottomrule
\end{tabular}
\vspace{-4mm}
\end{table}


The hardware synthesis results in Table~\ref{tab:implementation_overhead} show that FPGA implementation adds 6 DSPs with $\sim$11.8\% higher dynamic power over the base core. ASIC results (Synopsys Design Compiler v2024.09, typical corner, default switching activity) indicate 24.7\% area and 5.1\% power overhead. To evaluate area–delay efficiency, we report normalized Area-Delay Product (ADP) with respect to the baseline, where delay is measured using unpruned ViT-Base execution time; lower ADP implies better overall efficiency. As shown, the extended core achieves a lower ADP than the baseline.


\section{Conclusion}
This work presents an end-to-end framework for runtime-adaptive Vision Transformer (ViT) inference on embedded RISC-V processors. We combine post-training structured pruning with a hardware-aligned group-of-8 scheme and mask-based skip logic, enabling a single compiled binary to support multiple sparsity levels without recompilation. To further improve efficiency, we introduce a custom \texttt{MAC2} ISA extension that exploits input reuse pattern in dominant linear projection kernels. On the target RISC-V, the proposed system achieves up to $2.8\times$ inference speedup, 33\% energy reduction, and up to $4.86\times$ storage savings. While there is an inherent trade-off between accuracy and efficiency across pruning levels, the proposed framework enables low-overhead switching between operating points during inference based on application requirements. Overall, aligning structured sparsity with hardware execution granularity, together with lightweight ISA support, enables efficient and flexible ViT deployment on edge devices. Future work will extend mask-based skipping to nonlinear layers and explore heterogeneous acceleration using vector or matrix compute engines.

\section*{Acknowledgment}
The authors would like to thank Prof. Lizy K. John (University of Texas at Austin) for the valuable feedbacks and insightful discussions on this work.
\footnotesize
\bibliographystyle{IEEEtran}
\bibliography{main.bib}

\end{document}